\documentclass[aps,prd,twocolumn,groupedaddress,nofootinbib,showpacs]{revtex4}

\usepackage{euscript,graphicx,amssymb,subfigure}

\usepackage{amsfonts}
\usepackage{amsmath}
\usepackage{amssymb}
\usepackage{graphicx}
\usepackage{euscript}
\usepackage{color}
\usepackage{subfigure}
\usepackage{hyperref}

\newcommand{\goodgap}{\hspace{\subfigtopskip} \hspace{\subfigbottomskip}}

\begin{document}

\title{Constraining massive gravity with recent cosmological data}

\author{Vincenzo F. Cardone}
\email{winnyenodrac@gmail.com}
\affiliation{I.N.A.F.\,-\,Osservatorio Astronomico di Roma, via Frascati 33, 00040 - Monte Porzio Catone (Roma), Italy}

\author{Ninfa Radicella}
\email{ninfa.radicella@uab.cat}
\affiliation{Departamento de F\'isica, Universitat Aut\`onoma de Barcelona, 08193 Bellaterra (Barcelona), Spain}

\author{Luca Parisi}
\email{parisi@sa.infn.it}
\affiliation{Dipartimento di Fisica ``E.R. Caianiello", Universit\`{a} di Salerno and I.N.F.N. - Sez. di Napoli - GC di Salerno, Via Ponte Don Melillo, 84084 Fisciano (Sa), Italy}

\begin{abstract}

A covariant formulation of a theory with a massive graviton and no negative energy state has been recently proposed as an alternative to the usual General Relativity framework. For a spatially flat homogenous and isotropic universe, the theory introduces modified Friedmann equations where the standard matter term is supplemented by four effective fluids mimicking dust, cosmological constant, quintessence and stiff matter, respectively. We test the viability of this massive gravity formulation by contrasting its theoretical prediction to the Hubble diagram as traced by Type Ia Supernovae (SNeIa) and Gamma Ray Bursts (GRBs), the $H(z)$ measurements from passively evolving galaxies, Baryon Acoustic Oscillations (BAOs) from galaxy surveys and the distance priors from the Cosmic Microwave Background Radiation (CMBR) anisotropy spectrum. It turns out that the model is indeed able to very well fit this large dataset thus offering a viable alternative to the usual dark energy framework. We finally set stringent constraints on its parameters also narrowing down the allowed range for the graviton mass.

\end{abstract}


\maketitle

\section{Introduction}

There is a limited but crucial number of cosmological evidences which does not fit in the scheme constituted by General Relativity (GR) and the Standard Model of particle physics. On the one hand, it seems that the amount of baryonic matter in galaxies and galaxy clusters is not sufficient to cause the observed behaviour of the gravitational field on those scales. On the other hand, recent data on the SNeIa Hubble diagram provide evidences for an accelerated expansion, also confirmed by several other independent cosmological observations \cite{SNeIaNobel,others}. Such a large dataset can be very well reproduced by the concordance $\Lambda$CDM model made out of cold dark matter (CDM) and cosmological constant $\Lambda$. Notwithstanding this remarkable success, the $\Lambda$CDM scenario is theoretically unappealing since it comprises two exotic matter sources, namely dark energy and dark matter, and is moreover plagued by several well known shortcomings.

A different constructive point of view is to address these problems introducing modifications of GR over large distances in order to possibly explain our ignorance about the 95\% of the energy and matter content of the Universe. In this setting, a massive deformation of GR is a plausible modified theory of gravity that is both phenomenologically and theoretically intriguing.

From the theoretical point of view, a small nonvanishing graviton mass is an open issue. The idea was originally introduced in the work of Fierz and Pauli \cite{fierz39}, who constructed a massive theory of gravity in a flat background that is ghost\,-\,free at the linearized level. Since then, a great effort has been put in extending the result at the nonlinear level and constructing a consistent theory. As a somewhat unanticipated result, we nowadays recognize that such models can be relevant for cosmology too as possible alternative candidates to drive the accelerated expansion without the need of any exotic component.

Here we study cosmological solutions in the framework of a covariant massive gravity model, recently proposed in \cite{derham10, chamseddine11}. At the linearized order, the mass term breaks the gauge invariance of GR. Moreover, in order to construct a consistent theory, nonlinear terms should be tuned to remove order by order the negative energy state in the spectrum \cite{boulware72}. The theoretical model under investigation follows from a procedure originally outlined in \cite{arkani03,creminelli05} and has been found not to show ghosts at least up to quartic order in the nonlinearities \cite{derahm11,hassan11}.

We focus on cosmological equations of such theory, considering a spatially flat Robertson\,-\,Walker (RW) model in presence of both matter and radiation. Due to the potential term, four new terms are present in the Friedmann equations mimicking dust, cosmological constant, quintessence and stiff matter. The presence of the cosmological constant and quintessence\,-\,like terms easily suggest the possibility to achieve an accelerated expansion, while the stiff matter one can guarantee the halting of the cosmic speed up in the intermediate redshift regime thus recovering the standard decelerating matter dominated epoch. In order to check whether this is indeed the case, we contrast the model predictions with a wide dataset thus also being able to constrain its parameters. As a side effect, this will also allow us to narrow down the range for the graviton mass.

The plan of the paper is as follows. Sect.\,II presents the basics of the theory and put down the modified Friedmann equations describing the background cosmic evolution. The data we use and some subtleties specific to the massive gravity framework are discussed in Sect.\,III, while the results of the likelihood analysis and the constraints on the massive gravity coupling quantities are given in Sect.\,IV. We finally summarize and highlight some important issues in the concluding Sect.\,V.

\section{Massive Gravity}

Let us consider a four dimensional manifold equipped with both a dynamical metric $g_{\mu \nu}$ and a non\,-\,dynamical flat metric $f_{\mu \nu}$. In order to determine the dynamics on this structure, the starting point is to consider the following action\footnote{Unless otherwise specified, we will set the speed of light $c = 1$.} \cite{N11}

\begin{equation}
{\cal{S}} = - \frac{1}{8 \pi G} \int{\left ( \frac{1}{2} R + m^2 {\cal{U}} \right ) d^4x} + {\cal{S}}_M
\label{totaction}
\end{equation}
where $G$ is the Newton coupling constant, $R$ the Ricci scalar for $g_{\mu\nu}$, ${\cal{U}}$ the potential, and ${\cal{S}}_M$ describes ordinary matter which is supposed to directly interact only with $g_{\mu\nu}$. The potential term, coupled through the graviton mass $m$, is given by

\begin{eqnarray}
{\cal{U}} & = & \frac{1}{2} \left ( {\cal{K}}^2  - {\cal{K}}^{\nu}_{\mu} {\cal{K}}^{\mu}_{\nu} \right )
+ \frac{c_3}{3!} \epsilon_{\mu \nu \rho \sigma} \epsilon^{\alpha \beta \gamma \sigma} {\cal{K}}_{\alpha}^{\mu} {\cal{K}}_{\beta}^{\nu} {\cal{K}}_{\gamma}^{\rho} \nonumber \\
~ & + & \frac{c_4}{4!} \epsilon_{\mu \nu \rho \sigma} \epsilon^{\alpha \beta \gamma \delta} {\cal{K}}_{\alpha}^{\mu} {\cal{K}}_{\beta}^{\nu} {\cal{K}}_{\gamma}^{\rho} {\cal{K}}_{\delta}^{\sigma} \ ,
\label{potential}
\end{eqnarray}
with $\epsilon_{\mu\nu\rho\sigma}$ the Levi\,-\,Civita symbol, while the field ${\cal{K}}$ is related to the flat metric $f_{\mu \nu}$ through

\begin{displaymath}
{\cal{K}}^{\mu}_{\nu} = \delta^{\mu}_{\nu} - \gamma^{\mu}_{\nu} \ ,
\end{displaymath}
and we have defined
\begin{displaymath}
\gamma^{\mu}_{\sigma} \gamma^{\sigma}_{\nu} = g^{\mu \sigma} f_{\sigma \nu} \ .
\end{displaymath}
The theory is fully assigned by the graviton mass $m_g = \hbar m/c$ and the two coupling parameters $(c_3, c_4)$.

In order to derive the cosmological equations, one has to insert the RW metric into the field equations obtained by the usual variational approach to the action (\ref{totaction}) and impose the constraints dictated by the Bianchi identities \cite{chamseddine11}. For generic values of the $(c_3, c_4)$ couplings and a spatially flat universe consistent with CMBR data \cite{WMAP7}, the expansion rate turns out to be \cite{volkov}

\begin{eqnarray}
\frac{3 H^2}{8 \pi G} & = & \rho_M(a) + \rho_{r}(a) \nonumber \\
~ & + & \tilde{m}^2 (4 c_3 + c_4 - 6) \nonumber \\
~ & + & \frac{3 \tilde{m}^2 \beta (3 - 3 c_3 - c_4)}{a} \nonumber \\
~ & + & \frac{3 \tilde{m}^2 \beta^2 (c_4 + 2 c_3 - 1)}{a^2} \nonumber \\
~ & - & \frac{  \tilde{m}^2 \beta^3 (c_3 + c_4)}{a^3} \ ,
\label{eq: mgfried}
\end{eqnarray}
where $a$ is the scale factor, $H = \dot{a}/a$ the usual Hubble parameter (with the dot denoting derivative with respect to cosmic time), $\tilde{m}^2 = (m^2 c^2)/(8 \pi G)$ and we have assumed a source term made out by dust ($M$) and radiation ($r$). Since matter is assumed to be minimally coupled to gravity only, its conservation equation will be the usual

\begin{equation}
\dot{\rho_i} + 3 H (\rho_i + p_i) = 0
\label{eq: cons}
\end{equation}
with $p_i = 0$ ($p_i = \rho_{r}/3)$ for $i = M$ $(i = r)$. Introducing the redshift $z = 1/a - 1$, we can therefore conveniently rewrite Eq.(\ref{eq: mgfried}) as

\begin{eqnarray}
E^2(z) & = & \Omega_{\Lambda}^{eff} + \Omega_Q^{eff} (1 + z) + \Omega_{S}^{eff} (1 + z)^2 \nonumber \\
~ & + & \Omega_M^{eff} (1 + z)^3 + \Omega_{r} (1 + z)^4
\label{eq: hubbleend}
\end{eqnarray}
with $E(z) = H(z)/H_0$ (the label $0$ denoting present day values), $\Omega_i$ the standard present day density parameter for the component $i$, while we have defined the effective density parameters for the massive gravity terms as

\begin{equation}
\left \{
\begin{array}{l}
\displaystyle{\Omega_{\Lambda}^{eff} = \tilde{\mu}^2 (4 c_3 + c_4 - 6)} \\
~ \\
\displaystyle{\Omega_{Q}^{eff} = 3 \tilde{\mu}^2 \beta (3 - 3 c_3 - c_4)} \\
~ \\
\displaystyle{\Omega_{S}^{eff} = 3 \tilde{\mu}^2 \beta^2 (c_4 + 2 c_3 - 1)} \\
~ \\
\displaystyle{\Omega_{M}^{eff} = \Omega_M - \tilde{m}^2 \beta^3 (c_3 + c_4)} \\
\end{array}
\right . \ ,
\label{eq: omgeff}
\end{equation}
with $\tilde{\mu}^2 = \tilde{m}^2/\rho_{crit}$ and $\rho_{crit} = 3 H_0^2/8 \pi G$ the present day critical density\footnote{Note that, since $m = c m_g/\hbar$, it has the dimension of  $(\text{length})^{-1}$ so that $\tilde{\mu}^2 = (m^2 c^2)/(3 H_0^2)$ is dimensionless.}. Note that the $(c_3, c_4, \beta)$ parameters entering Eq.(\ref{eq: mgfried}) are all dimensionless.

A look at Eq.(\ref{eq: hubbleend}) shows that a massive graviton has a double effect on the Hubble rate. First, as is somewhat expected, it changes the matter content shifting the dust density parameter from the usual $\Omega_M$ value to the effective one $\Omega_M^{eff}$. It is, however, worth stressing that, since the sign of $c_3 + c_4$ is not set from the theory, it is also possible that the effective matter content is smaller than the actual one. Actually, the most interesting feature is the presence of the three further terms $(\Omega_{\Lambda}^{eff}, \Omega_Q^{eff}, \Omega_S^{eff})$ mimicking a cosmological constant, a quintessence\,-\,like field and stiff matter, respectively. Depending on the values of $(c_3, c_4)$ and the graviton mass, it is therefore possible to reproduce the expected series of radiation domination, matter era and accelerating expansion (driven by both $\Omega_{\Lambda}^{eff}$ and $\Omega_Q^{eff}$) with only a small contribute from the unusual stiff matter term. It is this particular feature that makes massive gravity so appealing from a cosmological point of view motivating the present analysis.

In order to constrain the model parameters, it is actually more convenient to replace the effective density parameters with $w_0 = w_{eff}(z = 0)$ and $w_p = dw_{eff}/dz|_{z = 0}$ with $w_{eff}$ the equation of state (EoS) of the effective dark energy model having the same Hubble parameter than the massive gravity one. This quantity is defined as

\begin{eqnarray}
w_{eff}(z) & = & -1 + \left [ \frac{2}{3} \frac{d\ln{E(z)}}{d\ln{(1 + z)}} \right .\nonumber \\
~ & - & \left . \frac{\Omega_M^{eff} (1 + z)^3}{E^2(z)} - \frac{\Omega_{r} (1 + z)^4}{E^2(z)} \right ] \nonumber \\
~ & \times & \left [ 1 - \frac{\Omega_M^{eff} (1 + z)^3}{E^2(z)} - \frac{\Omega_{r} (1 + z)^4}{E^2(z)} \right ]^{-1}
\label{eq: weff}
\end{eqnarray}
so that, by using Eq.(\ref{eq: hubbleend}), we get

\begin{eqnarray}
w_{eff}(z) & = & - \left \{ 3 \left [ \Omega_{\Lambda}^{eff} + \Omega_Q^{eff} (1 + z) + \Omega_S^{eff} (1 + z)^2 \right ] \right \}^{-1} \nonumber \\
~ & \times & \left [ 3 \Omega_{\Lambda}^{eff} + 2 \Omega_Q^{eff} (1 + z) \right . \nonumber \\
~ & + & \left . \Omega_S^{eff} (1 + z)^2 - \Omega_{r} (1 + z)^4 \right ] \ .
\label{eq: weffmg}
\end{eqnarray}
Imposing now $E^2(z = 0) = 1$, $w_0 = w_{eff}(z = 0)$ and $w_p = dw_{eff}/dz|_{z = 0}$, it is only a matter of algebra to get

\begin{eqnarray}
\Omega_{\Lambda}^{eff} & = & \frac{9}{2} (1 - \Omega_M^{eff} - \Omega_{r}) w_0^2 \nonumber \\
~ & + & \frac{3}{2} (3 - 3 \Omega_M^{eff} - 4 \Omega_{r}) w_0 \nonumber \\
~ & + & \frac{3}{2} (1 - \Omega_M^{eff} - \Omega_{r}) w_p \nonumber \\
~ & + & (1 - \Omega_M^{eff} - 3 \Omega_{r}) \ ,
\label{eq: omleff}
\end{eqnarray}

\begin{eqnarray}
\Omega_S^{eff} & = & \frac{9}{2} (1 - \Omega_M^{eff} - \Omega_{r}) w_0^2 \nonumber \\
~ & + & \frac{3}{2} (5 - 5 \Omega_M^{eff} - 6 \Omega_{r}) w_0 \nonumber \\
~ & + & \frac{3}{2} (1 - \Omega_M^{eff} - \Omega_{r}) w_p \nonumber \\
~ & + & 3 (1 - \Omega_M^{eff} - 2 \Omega_{r}) \ ,
\label{eq: omseff}
\end{eqnarray}

\begin{equation}
\Omega_Q^{eff} = 1 - \Omega_{\Lambda}^{eff} - \Omega_S^{eff} - \Omega_M^{eff} - \Omega_{r} \ .
\label{eq: omqeff}
\end{equation}
Using these relations, we can conveniently parameterize the model in terms of the effective matter density parameter $\Omega_M^{eff}$ and the present day values of the effective EoS and its derivative $(w_0, w_p)$. It is worth stressing that, depending on $(\Omega_M^{eff}, \Omega_{\gamma}, w_0, w_p)$ combinations, it is possible that some of the $(\Omega_{\Lambda}^{eff}, \Omega_Q^{eff}, \Omega_S^{eff})$ parameters take a negative value. This is, however, not a problem since $(\Omega_{\Lambda}^{eff}, \Omega_Q^{eff}, \Omega_S^{eff})$ refer to effective fluids, not actual ones. This can also be seen from Eqs.(\ref{eq: omgeff}) which show that negative $\Omega_i^{eff}$ can indeed be obtained depending on the values of the massive gravity couplings $(c_3, c_4)$.

\section{Massive gravity vs data}

Eq.(\ref{eq: weffmg}) shows that the massive gravity introduces an effective dark energy fluid whose EoS has the right behaviour to drive cosmic acceleration. Indeed, should the $\Omega_{\Lambda}^{eff}$ term be the dominant at low $z$, we get $w_{eff}(z << 1) \simeq -1$ so that a $\Lambda$CDM\,-\,like expansion is obtained in the late universe. As $z$ increases, we can still have an accelerated expansion as far as $\Omega_Q^{eff}$ dominates leading to $w_{eff}(z) \simeq -2/3$ in the intermediate redshift regime, while the transition to a decelerating epoch is guaranteed by the stiff matter term $\Omega_S^{eff}$. These encouraging features have to substantiated by the comparison with the available data which can both validate the model and constrain its parameters.

Although the dataset we are going to use is a typical one, there are some subtleties specific of the massive gravity framework which require some caution. We therefore prefer to spend some words to describe how we compare the model to the each kind of dataset.

\subsection{Hubble diagram}

A first step in testing any proposed model is the comparison with the Hubble diagram, i.e. the distance modulus $\mu$ as a function of the redshift $z$. This is related to the Hubble parameter $E(z)$ as

\begin{equation}
\mu = 25 + 5 \log{d_L(z, {\bf p})}
\label{eq: defmu}
\end{equation}
with

\begin{equation}
d_L(z, {\bf p}) = (1 + z) r(z, {\bf p})
\label{eq: defdl}
\end{equation}
the luminosity distance and

\begin{equation}
r(z, {\bf p}) = \frac{c}{H_0} \int_{0}^{z}{\frac{dz^{\prime}}{E(z^{\prime}, {\bf p})}}
\label{eq: defrz}
\end{equation}
the comoving distance. Note that we have collectively denoted with ${\bf p}$ the set of model parameters.

A classical tracer of the Hubble diagram is represented by SNeIa so that we use the Union2 dataset \cite{Union2} comprising ${\cal{N}}_{SNeIa} = 557$ objects tracing the Hubble diagram over the redshift range $(0.015, 1.4)$. The corresponding likelihood term will read\,:

\begin{eqnarray}
\label{eq: deflikesneia}
{\cal{L}}_{SNeIa} & = & \frac{1}{(2 \pi)^{{\cal{N}}_{SNeIa}} | {\bf C}_{SNeIa} |^{1/2}} \\
~ & \times & \exp{\left [ - \frac{{\bf D}_{SNeIa}^{T}({\bf p}) {\bf C}_{SNeIa}^{-1} {\bf D}_{SNeIa}({\bf p})}{2} \right ]} \nonumber
\end{eqnarray}
with ${\bf C}_{SNeIa}$ the SNeIa covariance matrix and ${\bf D}({\bf p})$ a ${\cal{N}}_{SNeIa}$\,-\,dimensional vector with the $i$\,-\,th element given by $\mu_{obs}(z_i) - \mu_{th}(z_i, {\bf p})$, i.e. the difference between the observed and predicted distance modulus. Neglecting systematics, the covariance matrix becomes diagonal and we can simplify Eq.(\ref{eq: deflikesneia}) as

\begin{equation}
{\cal{L}}_{SNeIa} = \frac{1}{(2 \pi)^{{\cal{N}}_{SNeIa}}\Gamma_{SNeIa}^{1/2}} \
\exp{\left [ - \frac{\chi^2_{SNeIa}({\bf p})}{2} \right ]}
\label{eq: deflikesneiabis}
\end{equation}
where

\begin{equation}
\chi^2_{SNeIa} = \sum_{i = 1}^{{\cal{N}}_{SNeIa}}{\left [ \frac{\mu_{obs}(z_i) - \mu_{th}(z_i, {\bf p})}{\sigma_{i}} \right ]^2} \ ,
\label{eq: defchisneia}
\end{equation}

\begin{equation}
\Gamma_{SNeIa} = \prod_{i = 1}^{{\cal{N}}_{SNeIa}}{\sigma_{i}^2} \ .
\label{eq: defgammasneia}
\end{equation}
While SNeIa efficiently probe the late universe thus being highly sensible to the present day cosmic speed up, the transition to the decelerated expansion and the matter dominated era are better investigated resorting to higher $z$ tracer. This is provided by GRBs so that we rely on their Hubble diagram as derived in \cite{Marcy} based on a model independent calibration of five different scaling relations and the catalog given in \cite{XS10}. Note that we cut the sample only using ${\cal{N}}_{GRB} = 64$ objects probing the range $1.48 \le z \le 5.60$ in order to avoid any residual correlations with the SNeIa sample (which is used to calibrate the GRBs scaling relations). The GRB likelihood term is similar to the SNeIa being defined as

\begin{equation}
{\cal{L}}_{GRB} = \frac{1}{(2 \pi)^{{\cal{N}}_{GRB}}\Gamma_{GRB}^{1/2}} \
\exp{\left [ - \frac{\chi^2_{GRB}({\bf p})}{2} \right ]}
\label{eq: deflikegrb}
\end{equation}
where

\begin{equation}
\chi^2_{GRB} = \sum_{i = 1}^{{\cal{N}}_{GRB}}{\left [ \frac{\mu_{obs}(z_i) - \mu_{th}(z_i, {\bf p})}{\sqrt{\sigma_{i}^2 + \sigma_{int}^2}} \right ]^2} \ ,
\label{eq: defchigrb}
\end{equation}

\begin{equation}
\Gamma_{GRB} = \prod_{i = 1}^{{\cal{N}}_{GRB}}{\left ( \sigma_{i}^2 + \sigma_{int}^2 \right )} \ .
\label{eq: defgammagrb}
\end{equation}
Note that we have here added to the measurement uncertainties $\sigma_i$ the intrinsic scatter $\sigma_{int}$ which takes into account the dispersion of the single GRBs around the input scaling relations. We will marginalize over $\sigma_{int}$ so that we will not give any constraint on this quantity.

\subsection{$H(z)$ data}

Being a probe of the integrated Hubble parameter, the SNeIa\,+\,GRB Hubble diagram smooths out the deviations from the (unknown) best fit model provided these are not too extreme. It is therefore desirable to add a second dataset which directly probes the $E(z)$ which is indeed possible resorting to the differential age method \cite{JL02}. Such a technique is motivated by noting that, from the relation $dt/dz = -(1+z) H(z)$, a measurement of $dt/dz$ at different $z$ gives, in principle, the Hubble parameter. Good results can be obtained by using fair samples of passively evolving galaxies with similar metallicity and low star formation rate so that they can be taken to be the oldest objects at a given $z$. Stern et al. \cite{S10II} have then used red envelope galaxies as cosmic chronometers determining their ages from high quality Keck spectra and applied the differential age method to estimate $H(z)$ over the redshift range $0.10 \le z \le 1.75$ \cite{S10I}. We use their data as input to the following likelihood function\,:

\begin{equation}
{\cal{L}}_H({\bf p}) = \frac{1}{(2 \pi)^{{\cal{N}}_H/2} \Gamma_H^{1/2}} \ \exp{\left [ - \frac{\chi_H^2({\bf p})}{2} \right ]} \ ,
\label{eq: deflikehz}
\end{equation}
with

\begin{equation}
\chi^2_H = \sum_{i = 1}^{{\cal{N}}_H}{\left [ \frac{H_{obs}(z_i) - H(z_i, {\bf p})}{\sigma_{hi}} \right ]^2} \ ,
\label{eq: defchihz}
\end{equation}

\begin{equation}
\Gamma_{H} = \prod_{i = 1}^{{\cal{N}}_{H}}{\sigma_{hi}^2} \ .
\label{eq: defgammahz}
\end{equation}
where ${\cal{N}}_H = 11$ is the number of sample points.

\subsection{Baryon Acoustic Oscillations}

An alternative and very promising method to trace the distance\,-\,redshift relation relies on the measurement of baryon acoustic oscillations (BAOs) in the large scale clustering pattern of galaxies \cite{BAOth}. BAOs correspond to a preferred length scale imprinted in the distribution of photons and baryons by the propagation of sound waves in the relativistic plasma of the early universe. This length scale, corresponding to the sound horizon $r_s(z_d)$ at the baryon drag epoch, manifests itself in the clustering pattern of galaxies as a small preference for pairs of galaxies to be separated by $r_s(z_d)$, causing a distinctive peak in the 2\,-\,point galaxy correlation function. Since the Fourier transform of a $\delta$\,-\,function is a $sinc$, this signature will look like a series of decaying oscillations in the galaxy power spectrum thus motivating the BAO name. The small amplitude of the peak and the large size of the relevant scales imply that large volumes $(\sim 1 \ {\rm Gpc}^3)$ and a high number of galaxies $(\sim 10^5)$ must be observed in order to ensure a robust detection. It is therefore not surprising that the first reliable determination \cite{Eis05} has to await for the third data release of the SDSS survey with other determinations \cite{BAOobs} all relying on similarly large (both spectroscopic and photometric) galaxy surveys.

In order to include BAOs as constraint, we use the measurement at six different redshifts, namely $z = 0.106$ from 6dFGRS \cite{6dFGRS}, $z = (0.20, 0.35)$ from SDSS\,+\,2dFGRS \cite{P10}, and $z = (0.44, 0.60, 0.73)$ from WiggleZ \cite{WiggleZ}. Since the three surveys are independent on each other, we can define the likelihood function as

\begin{displaymath}
{\cal{L}}_{BAO} = {\cal{L}}_{6dFGRS} \ \times \ {\cal{L}}_{SDSS} \ \times \ {\cal{L}}_{WiggleZ}
\end{displaymath}
with the three terms given by \cite{WiggleZ}

\begin{equation}
{\cal{L}}_{6dFGRS} = \frac{1}{\sqrt{2 \pi \sigma_{0.106}^2}} \ \exp{\left \{ - \frac{1}{2}
\left [ \frac{d_{0.106}^{obs} - d_{0.106}^{th}({\bf p})}{\sigma_{0.106}} \right ]^2 \right \}} \ ,
\label{eq: deflike6df}
\end{equation}

\begin{eqnarray}
{\cal{L}}_{SDSS} & = & \frac{1}{(2 \pi)^{{\cal{N}}_{SDSS}} | {\bf C}_{SDSS} |^{1/2}} \\
~ & \times & \exp{\left [ - \frac{{\bf D}_{SDSS}^{T}({\bf p}) {\bf C}_{SDSS}^{-1} {\bf D}_{SDSS}({\bf p})}{2} \right ]} \nonumber \ ,
\label{eq: deflikesdss}
\end{eqnarray}

\begin{eqnarray}
{\cal{L}}_{WiggleZ} & = & \frac{1}{(2 \pi)^{{\cal{N}}_{WiggleZ}} | {\bf C}_{WiggleZ} |^{1/2}} \\
~ & \times & \exp{\left [ - \frac{{\bf D}_{WiggleZ}^{T}({\bf p}) {\bf C}_{WiggleZ}^{-1} {\bf D}_{WiggleZ}({\bf p})}{2} \right ]} \nonumber \ .
\label{eq: deflikewigglez}
\end{eqnarray}
In Eq.(\ref{eq: deflikesdss}), $D_{SDSS}$ is a two dimensional vector with the values of $d_z^{obs} - d_z^{th}({\bf p})$ having defined

\begin{equation}
d_z = \frac{r_s(z_d)}{d_V(z)} = r_s(z_d) \times \left [ \frac{c z r^2(z)}{H_0 E(z)} \right ]^{-\frac{1}{3}} \ ,
\label{eq: defdz}
\end{equation}
with the sound horizon to distance $z$ given by

\begin{equation}
r_s(z) = \frac{c}{\sqrt{3} H_0} \int_{z}^{\infty}{\frac{E^{-1}(z) dz^{\prime}}{\sqrt{1 + (3 \omega_b)/(4 \omega_r) (1 + z^{\prime})^{-1}}}} \ .
\label{eq: defrs}
\end{equation}
Note that we have here introduced the physical baryon and photon density parameters defined as $\omega_i = \Omega_i h^2$ (with $h$ the Hubble constant in units of $100 \ {\rm km/s/Mpc}$), while $z_d$ is the drag epoch redshift. We will follow \cite{6dFGRS} to set $(d_{0.106}^{obs}, \sigma_{0.106})$, while \cite{P10} gives the relevant $d_z^{obs}$ values and covariance matrix for the measurement at $z = (0.20, 0.35)$ from SDSS\,+\,2dFGRS data.

The WiggleZ survey has recommended to use a different BAO related quantity so that $D_{WiggleZ}$ is a three dimensional vector whose $i$\,-\,th element is given by the difference between the observed and predicted value of the acoustic parameter ${\cal{A}}(z)$ defined as \cite{Eis05,WiggleZ}

\begin{equation}
{\cal{A}}(z) = \frac{\sqrt{\Omega_M H_0^2} d_V(z)}{c z}
\label{eq: defaz}
\end{equation}
with the volume distance $d_V(z)$ given in Eq.(\ref{eq: defdz}). We use the observed values and their covariance matrix for ${\cal{A}}(z)$ determinations at $z = (0.44, 0.60, 0.70)$ reported in \cite{WiggleZ}.

An important remark is in order here. Both the drag epoch redshift $z_d$ and the acoustic parameter ${\cal{A}}(z)$ explicitly depend on the matter physical density parameter $\omega_M$. It is worth wondering which value has to be used for $\Omega_M$ in the definition of the $\omega_M$, i.e. whether it is $\omega_M = \Omega_M h^2$ or $\omega_M = \Omega_M^{eff} h^2$. This is a subtle question that has  not a definitive answer at the moment. Indeed, both $z_d$ and ${\cal{A}}(z)$ are related to the clustering properties of the matter so that one should investigate how clustering takes place in massive gravity to understand if it depends on the amount of either effective or actual matter. In order to be conservative, we have decided to set $\omega_M = \Omega_M h^2$ when computing both $z_d$ and ${\cal{A}}(z)$ since the standard dust matter is the only one which surely clusters. Actually, since the graviton mass is expected to be very small, we can anticipate that $\Omega_M$ and $\Omega_M^{eff}$ will not be very different so that, should our choice be the incorrect one, the estimated $z_d$ and ${\cal{A}}(z)$ values will be biased low by a very small amount.

\subsection{CMBR data}

Combining the data described above, we are able to trace the background expansion of the late and intermediate $z$ universe. CMBR data, on the contrary, give us a picture of the universe in its infancy, at the redshift of the last scattering surface ($z_{\star} \sim 1100$). As shown in \cite{WMAP7}, rather than fitting the full CMBR anisotropy spectrum, one can get reliable and almost equivalent constraints by considering the so called distance priors, namely the redshift $z_{\star}$ of the last scattering surface, the acoustic scale $\ell_A = \pi r(z_{\star})/r_s(z_{\star})$ and the shift parameter ${\cal{R}}$ \cite{ShiftPar}

\begin{equation}
{\cal{R}} = \frac{\sqrt{\Omega_M^{eff}} r(z_{\star})}{c/H_0} \ .
\label{eq: defshiftpar}
\end{equation}
The CMBR likelihood function is then defined as

\begin{eqnarray}
{\cal{L}}_{CMBR} & = & \frac{1}{(2 \pi)^{{\cal{N}}_{CMBR}} | {\bf C}_{CMBR} |^{1/2}} \\
~ & \times & \exp{\left [ - \frac{{\bf D}_{CMBR}^{T}({\bf p}) {\bf C}_{CMBR}^{-1} {\bf D}_{CMBR}({\bf p})}{2} \right ]} \nonumber \ ,
\label{eq: deflikecmbr}
\end{eqnarray}
where the ${\cal{N}}_{CMBR} (= 3)$-dimensional vector $D_{CMBR}$ contains the difference among observed and theoretically predicted values of $(\ell_A, {\cal{R}}, z_{\star})$. We take the observed distance priors and the corresponding covariance matrix from \cite{WMAP7} and use the approximated formula in \cite{HS96} to compute $z_{\star}$ as function of $(\omega_b, \omega_M)$. We stress that we set here $\omega_M = \Omega_M h^2$, but use $\Omega_M^{eff}$ in the ${\cal{R}}$ estimate. This latter choice has been motivated by the consideration that the shift parameter actually turns out by neglecting terms other than the $(1 + z)^3$ in the evaluation of the distance to the last scattering surface. Since the importance of this term is determined by the $\Omega_M^{eff}$ parameter, it is this quantity which must enter the ${\cal{R}}$ definition.

\subsection{The full likelihood function}

Since the single datasets are independent on each other, combining all of them in a single fit is straightforward. One has only to define a full likelihood as

\begin{figure*}
\centering
\subfigure{\includegraphics[width=5.0cm]{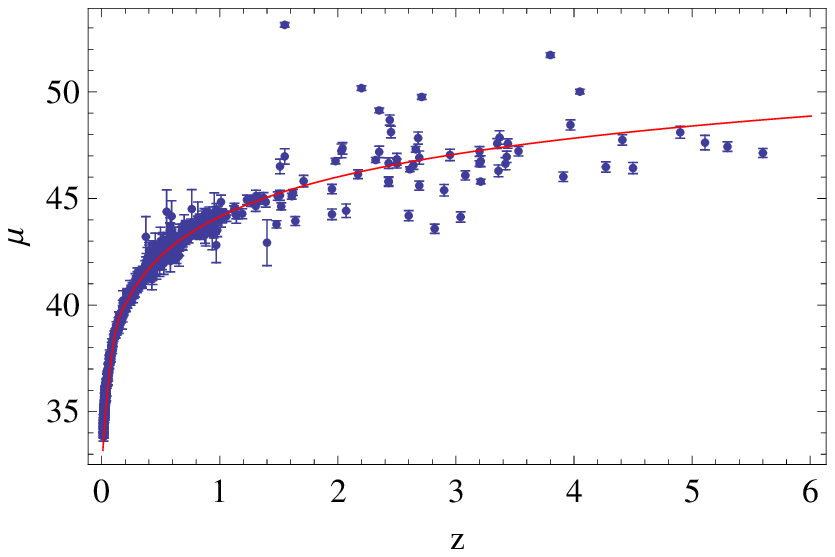}} \goodgap
\subfigure{\includegraphics[width=5.0cm]{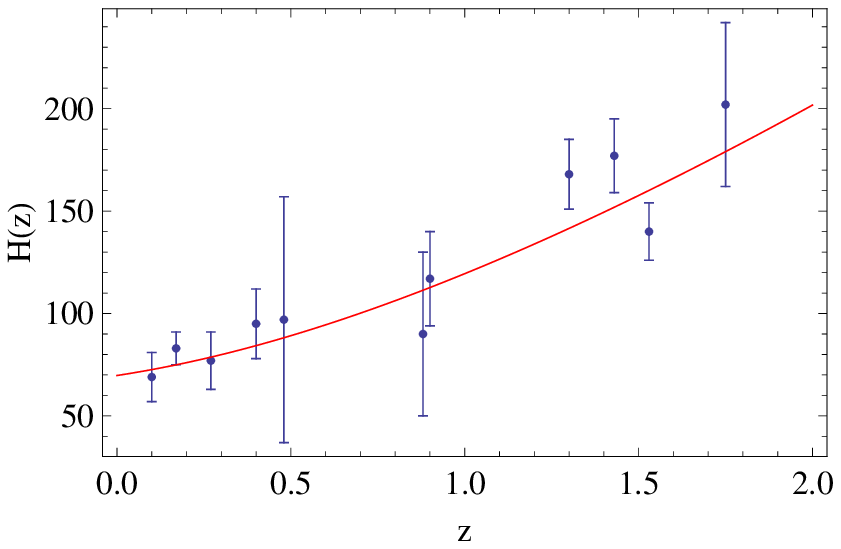}} \goodgap
\subfigure{\includegraphics[width=5.0cm]{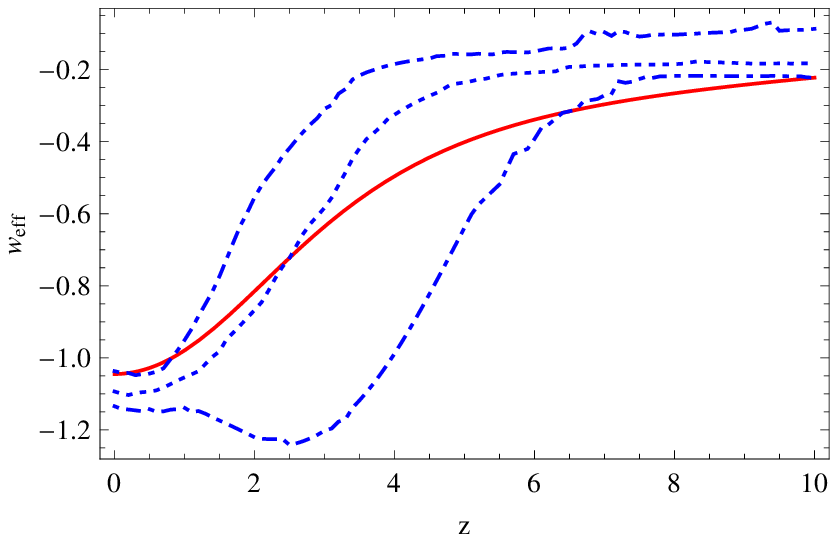}} \goodgap
\caption{{\it Left.} Best fit model superimposed to the Hubble diagram. {\it Centre.} Best fit model overlapped with the $H(z)$ data. {\it Right.} Effective EoS with the solid line referring to the best fit model and the dashed ones to the median and $68\% \ {\rm CL}$.}
\label{fig: bfplot}
\end{figure*}

\begin{displaymath}
{\cal{L}}_{data} = {\cal{L}}_{SNeIa} \times {\cal{L}}_{GRB} \times {\cal{L}}_{H} \times {\cal{L}}_{BAO} \times {\cal{L}}_{CMBR} \times {\cal{L}}_{0}
\end{displaymath}
where the last term is a Gaussian prior on $h$ given by the SHOES collaboration \cite{SHOES} which have determined $h =~0.738 \pm 0.024$ from the local distance ladder.

According to the above review of the data, ${\cal{L}}_{data}$ will depend on the massive gravity parameters $(\Omega_M^{eff}, w_0, w_p)$, the present day Hubble constant $h$ and the physical density parameters $(\omega_b, \omega_M, \omega_{r})$ of baryons, matter and radiation. In order to reduce the parameter space, we will set $(\omega_b, \omega_{\gamma}) = (2.258 \times 10^{-2}, 2.469 \times 10^{-5})$ in agreement with \cite{WMAP7} and set the radiation density parameter as

\begin{displaymath}
\Omega_{r} = \omega_{\gamma} h^{-2} (1 + 0.2271 N_{eff})
\end{displaymath}
with $N_{eff} = 3.04$ the effective neutrino number.

Should we rely on ${\cal{L}}_{data}$ only, nothing would prevent the likelihood to favour models with reasonable values of $(\Omega_M^{eff}, w_0, w_p, h, \omega_M)$, but leading to unphysical massive gravity parameters. Indeed, while Eqs.(\ref{eq: omleff})\,-\,(\ref{eq: omqeff}) always give real values whatever the $(\Omega_M^{eff}, w_0, w_p)$ parameters are, solving Eqs.(\ref{eq: omgeff}) with respect to $(\tilde{\mu}^2, \beta, c_3, c_4)$ can lead to unphysical values such as a negative $\tilde{m}^2$. In order to avoid this possibility, we therefore define our final likelihood function as

\begin{equation}
{\cal{L}}({\bf p}) = {\cal{L}}_{data}({\bf p}) \ \times \ {\cal{P}}({\bf p})
\label{eq: endlike}
\end{equation}
with ${\cal{P}}({\bf p}) = 1$ if $i.)$ $\tilde{\mu}^2 > 0$, $ii.) \beta > 0$ (in order to avoid a change in the signature of the metric) and $iii.)$ $c_3 c_4 < 0$ \cite{K11}, while it is ${\cal{P}}({\bf p}) = 0$ otherwise.

In order to efficiently explore the parameter space, we use a Markov Chain Monte Carlo (MCMC) method running three parallel chains and keep adding points until the Gelman\,-\,Rubin \cite{GR92} convergence criterium is satisfied. The best fit model will be the one maximizing the likelihood function ${\cal{L}}({\bf p})$, but, in a Bayesian framework, most reliable constraints on a single parameter $p_j$ are given by the mean, median and $68\%$ and $95\%$ confidence levels (CL) estimated from the likelihood function after marginalizing over all the parameters but the $j$\,-\,th one. Actually, the marginalization is implemented by simply looking at the histograms of the merged chain after cutting the burn\,-\,in phase and thinning to avoid spurious correlations.

\section{Results}

Three parallel chains with $\sim 13000$ points each turn out to be sufficient to achieve a good convergence of the MCMC code, the $R$ parameter being smaller than $1.1$ for all the fitted parameters. The best fit model read

\begin{displaymath}
\Omega_M^{eff} = 0.282 \ \ , \ \ \omega_M = 0.1351 \ \ , \ \ h = 0.697 \ \ ,
\end{displaymath}

\begin{displaymath}
w_0 = -1.04 \ \ , \ \ w_p = 0.00 \ \ ,
\end{displaymath}
while its theoretical predictions for the Hubble diagram and the Hubble parameter are superimposed to the data in Fig.\,\ref{fig: bfplot}. Here, we also plot the effective dark energy EoS for the best fit model and the constraints from the MCMC analysis. To this end, for each $z$, we evaluate $w_{eff}(z)$ for all the points in the chain and infer the median value and $68\%$ confidence ranges from the resulting distribution. Note that the best fit $w_{eff}(z)$ and the median thus obtained are not equal because of the degeneracies in the model parameters space so that it is actually the median $w_{eff}(z)$ the most reliable one in a Bayesian framework since it takes explicitly into account such degeneracies and marginalize over them.

It is interesting to look in detail to how the best fit model compares to the observed data to guess some hint on pro and cons of the massive gravity framework. Indeed, while the low reduced $\chi^2$ values make us forecast that the agreement will be very good, it is nevertheless worth checking what happens for each single probe to see whether the likelihood is dominated by a single term only\footnote{To understand why this could happen, one can imagine to fit only the SNeIa and GRBs Hubble diagrams. Since there are an order of magnitude more SNeIa than GRBs, should the fit to the two separated datasets points towards different regions of the parameter space, the fit to the combined sample will be closer to the SNeIa only one. In such a case, one could still find a low reduced $\chi^2_{GRB}$ because of the larger uncertainties on this dataset. Actually, this does not take place, but we have preferred to explicitly check for this possibility.}. First, we note that the model very well fits the Hubble diagram data with

\begin{table}
\begin{center}
\begin{tabular}{cccccc}
\hline
Id & $x_{BF}$ & $\langle x \rangle$ & $\tilde{x}$ & $68\% \ {\rm CL}$ & $95\% \ {\rm CL}$ \\
\hline \hline
~ & ~ & ~ & ~ & ~ & ~ \\
$\Omega_M^{eff}$ & 0.282 & 0.284 & 0.284 & (0.272, 0.293) & (0.260, 0.306) \\
~ & ~ & ~ & ~ & ~ & ~ \\
$\omega_M$ & 0.1351 & 0.1375 & 0.1380 & (0.1331, 0.1412) & (0.1274, 0.1479) \\
~ & ~ & ~ & ~ & ~ & ~ \\
$h$ & 0.697 & 0.705 & 0.703 & (0.695, 0.714) & (0.689, 0.725) \\
~ & ~ & ~ & ~ & ~ & ~ \\
$w_0$ & -1.04 & -1.09 & -1.09 & (-1.13, -1.04) & (-1.17, -0.98) \\
~ & ~ & ~ & ~ & ~ & ~ \\
$w_p$ & 0.00 & -0.05 & -0.05 & (-0.09, -0.01) & (-0.11, 0.02) \\
~ & ~ & ~ & ~ & ~ & ~ \\
\hline
\end{tabular}
\caption{Constraints on the fit parameters. Columns are as follows\,: 1.) id parameter, 2.) best fit, 3.) mean, 4.) median, 5.), 6.) $68\%$ and $95\%$ confidence ranges.}
\label{tab: fitpar}
\end{center}
\end{table}

\begin{displaymath}
\chi^2_{SNeIa}/d.o.f. = 0.98 \ \ , \ \ \chi^2_{GRB}/d.o.f. = 1.08 \  \ ,
\end{displaymath}
$d.o.f. = {\cal{N}}_i - n_p$ being the number of degrees of freedom. This is a nice consequence of the effective EoS turning out to be very close to the $\Lambda$CDM one, which is parameterized by $(w_0, w_p) = (-1, 0)$. The smoothing process introduced by the integration entering the luminosity distance definition cancels out the residual differences which, on the contrary, are better appreciated when fitting the $H(z)$ data leading to $\chi^2_H/d.o.f. = 1.23$. This is still a fully acceptable result considering both the low statistics (giving $d.o.f. = 5$) and the scatter of the $H(z)$ data around the best fit $\Lambda$CDM prediction.

The model turns out to be well efficient in reproducing also the BAO data. For the distance ratios, we get

\begin{displaymath}
\left \{
\begin{array}{lll}
d_{0.106}^{bf} = 0.3435 & {\rm vs} & d_{0.106}^{obs} = 0.336 \pm 0.015 \\
~ & ~ & ~ \\
d_{0.200}^{bf} = 0.1872 & {\rm vs} & d_{0.200}^{obs} = 0.1905 \pm 0.0061 \\
~ & ~ & ~ \\
d_{0.350}^{bf} = 0.1123 & {\rm vs} & d_{0.350}^{obs} = 0.1097 \pm 0.0036 \\
\end{array}
\right . \ ,
\end{displaymath}
while the acoustic parameter values read

\begin{displaymath}
\left \{
\begin{array}{lll}
{\cal{A}}^{bf}(0.44) = 0.460 & {\rm vs} & {\cal{A}}^{obs}(0.44) = 0.474 \pm 0.034 \\
~ & ~ & ~ \\
{\cal{A}}^{bf}(0.60) = 0.435 & {\rm vs} & {\cal{A}}^{obs}(0.60) = 0.442 \pm 0.020 \\
~ & ~ & ~ \\
{\cal{A}}^{bf}(0.73) = 0.414 & {\rm vs} & {\cal{A}}^{obs}(0.73) = 0.424 \pm 0.021 \\
\end{array}
\right . \ .
\end{displaymath}
As it is apparent, the theoretically predicted values agree with the observed one well within $1 \sigma$ so that we can be well confident in the ability of the model to give the correct BAO features. It is worth noting, however, that the acoustic parameter values are systematically smaller than the observed one with ${\cal{A}}^{bf}(z)/{\cal{A}}^{obs}(z) \simeq 0.98$ independent on $z$. Noting that, for the best fit values, $\omega_M/\Omega_M^{eff} h^2 = 0.99$, one can argue that the observed discrepancy could be remedied by assuming that the full effective matter term enters the clustering process or, at least, should be taken into account when computing the BAO acoustic parameter at different redshifts.

Finally, we compare the theoretically predicted distance priors with the observed ones being

\begin{displaymath}
\left \{
\begin{array}{lll}
\ell_A^{bf} = 302.023 & {\rm vs} & \ell_A^{obs} = 302.09 \pm 0.76 \\
~ & ~ & ~ \\
{\cal{R}}^{bf} = 1.729 & {\rm vs} & {\cal{R}}^{obs} = 1.725 \pm 0.018 \\
~ & ~ & ~ \\
z_{\star}^{bf} = 1090.92 & {\rm vs} & z_{\star}^{obs} = 1091.30 \pm 0.91 \\
\end{array}
\right .
\end{displaymath}
so that the agreement is still excellent.

The above results have been obtained contrasting the model to the full observational probes we have described in Sect.\,III which also include data from GRBs and the $H(z)$ as inferred from galaxy ages. Although both these datasets are well fitted by the model, their use as cosmological probes is quite recent so that one can not exclude a priori that some undetected systematic is present and potentially bias the constraints. As a consistency check, we have therefore repeated the above analysis removing both GRBs and $H(z)$ data from the fit. The median value and $68\%$ CL turn out to be\,:

\begin{displaymath}
\Omega_M^{eff} = 0.282_{-0.012}^{+0.012} \ , \ \omega_M = 0.1316_{-0.0061}^{+0.0039} \ ,
\end{displaymath}

\begin{displaymath}
h = 0.685_{-0.008}^{+0.012} \ , \ w_0 = -1.06_{-0.05}^{+0.05} \ , \ w_p = 0.06_{-0.02}^{+0.03} \ .
\end{displaymath}
These results are in good agreement with those in Table\,\ref{tab: fitpar} with the $68\%$ CL well overlapping for all the parameters but $w_p$. This latter quantity is the one with the largest shit since now positive rather than negative values are preferred. Actually, for both fits, the results argue in favour of a very mild evolution of the effective EoS which can be easily smoothed out from the integration needed to get the luminosity distance. We therefore expect that adding $H(z)$ data erases this discrepancy. Indeed, fitting this dataset too (but still excluding GRBs), we get\,:

\begin{displaymath}
\Omega_M^{eff} = 0.284_{-0.006}^{+0.008} \ , \ \omega_M = 0.1399_{-0.0061}^{+0.0023} \ ,
\end{displaymath}

\begin{displaymath}
h = 0.710_{-0.010}^{+0.005} \ , \ w_0 = -1.07_{-0.05}^{+0.03} \ , \ w_p = -0.07_{-0.03}^{+0.03} \ ,
\end{displaymath}
which are now in still better agreement with those in Table\,\ref{tab: fitpar} with the $w_p$ confidence range moving towards negative values. We also note that the median $h$ value is larger and almost equal to the one from the fit to the full dataset. This comes out as a consequence of the need for the fit to {\it align} the present day $H_0$ value to the trend followed by the $z > 0$ measurements. That this is the case is also confirmed by the last test we do removing the $H(z)$ data, but adding the GRB ones. We now get\,:

\begin{displaymath}
\Omega_M^{eff} = 0.284_{-0.012}^{+0.011} \ , \ \omega_M = 0.1347_{-0.0055}^{+0.0048} \ ,
\end{displaymath}

\begin{displaymath}
h = 0.695_{-0.005}^{+0.010} \ , \ w_0 = -1.07_{-0.06}^{+0.04} \ , \ w_p = -0.01_{-0.03}^{+0.02} \ .
\end{displaymath}
We indeed find a smaller $h$ value, but still a good agreement with the results in Table\,\ref{tab: fitpar}. We therefore conclude that, should some undetected systematics be present in the GRBs and/or $H(z)$ data, the constraints would not be affected so that we can safely rely on them.

\begin{table}
\begin{center}
\begin{tabular}{cccccc}
\hline
Id & $x_{BF}$ & $\langle x \rangle$ & $\tilde{x}$ & $68\% \ {\rm CL}$ & $95\% \ {\rm CL}$ \\
\hline \hline
~ & ~ & ~ & ~ & ~ & ~ \\
$m_g$ & 2.85 & 7.26 & 0.06 & (0.02, 0.96) & (0.01, 30.6) \\
~ & ~ & ~ & ~ & ~ & ~ \\
$\beta$ & 0.39 & 0.38 & 0.41 & (0.16, 0.54) & (0.05, 0.62) \\
~ & ~ & ~ & ~ & ~ & ~ \\
$c_3$ & 2.42 & 5.91 & 3.86 & (1.57, 11.2) & (1.31, 21.2) \\
~ & ~ & ~ & ~ & ~ & ~ \\
$c_4$ & -2.88 & -8.15 & -5.52 & (-15.8, -0.82) & (-29.8, -0.70) \\
~ & ~ & ~ & ~ & ~ & ~ \\
\hline
\end{tabular}
\caption{Constraints on the massive gravity parameters with $m_g$ the graviton mass in $10^{-66} \ {\rm g}$ units.}
\label{tab: mgtab}
\end{center}
\end{table}

Having convincingly shown that the model is able to reproduce a wide set of data, it is worth looking at the constraints on the quantities explicitly entering the massive gravity Lagrangian. For each set of fitted parameters along the merged MCMC chain, we therefore solve for the $(\tilde{\mu}^2, \beta, c_3, c_4)$ and give the inferred constraints in Table\,\ref{tab: mgtab} where we report the graviton mass $m_g = 3 H_0 \tilde{\mu} \hbar/c^2$ instead of $\tilde{\mu}$ since the former is more interesting from a theoretical point of view.

A caveat is in order here. Solving for $(\tilde{\mu}^2, \beta, c_3, c_4)$ as function of the fitted parameters can give rise to large values of the $(c_3, c_4)$ couplings for particular combinations of the input quantities. As a consequence, the $(c_3, c_4)$ distributions turn out to be strongly asymmetric with $\sim 90\%$ of the points being close to the median value, but the remaining $\sim 10\%$ distributed in a very long tail towards extremal values (positive for $c_3$ and negative for $c_4$). A similar argument also applies for $m_g$ which has a long tail towards large masses. In order to avoid this effect, we have arbitrarily cut these tails rejecting the most extreme $10\%$ points and recomputing the constraints from the survival points. These are the values reported in Table\,\ref{tab: mgtab} which can be considered as a compromise between what data tell us and what is actually reasonable. It is worth stressing that the above cut has only a minor effect on the constraints on the graviton mass.

\section{Conclusions}

Until not long ago, it was thought to be impossible to construct a consistent theory of  massive gravity compatible with current observations \cite{creminelli05, deffayet05}. Only in 2010, de Rham and Gabadadze \cite{derham10} presented a good candidate for a ghost\,-\,free completion of the Fierz\,-\,Pauli theory by choosing interactions with no more than two time derivatives of the scalar degrees of freedom \cite{derham10}. The cosmological sector of this theory have been recently worked out in \cite{chamseddine11,volkov} thus setting the necessary framework to compare the model predictions with the observed universe. We have then been able to constrain the model parameters using a large dataset comprising the SNeIa\,+\,GRB Hubble diagram, $H(z)$ measurements from cosmic chronometers, BAOs data and the CMBR distance priors. It turns out that the model is in very good agreement with the data thus giving an observationally motivated support to massive gravity as a theoretically appealing alternative the the troublesome $\Lambda$CDM scenario.

Lacking up to now a clear physical interpretation of the $(c_3, c_4)$ couplings entering the modified gravity Lagrangian, one can only indirectly check which are the conditions they have to fulfil in order the theory be consistent. To this end, we stress that our constraints are consistent with the conditions $c_3 + c_4 = 0$ and $c_4 \neq -1$, but the case $c_3 + c_4 > 0$ is still allowed. In both cases, the constrained parameter space refers to models which have the right properties to recover GR in the low energy limit as demonstrated by \cite{koyama12} through the study of spherically symmetric solutions. We therefore end up with a massive gravity theory able to reproduce the cosmological data without altering the success of the standard General Relativity on the Solar System scales.

As a further relevant result, the comparison of the theory with the cosmological data has allowed us to set a limit on the graviton mass. Although an explicit derivation is needed, we can assume that, in the low energy limit, the gravitational potential will be boosted by a Yukawa\,-\,like correction modulated by the length scale $\lambda_g \propto 1/m_g$. One can then compare our results with the upper limit $m_g < 7.68 \times 10^{-55} \ {\rm g}$ from the dynamics in the Solar System \cite{T88} and the  more stringent limit, $m_g < 10^{-59} \ {\rm g}$, derived by requiring the derived dynamical properties of a galactic disk to be consistent with observations \cite{disk}. Our $95\%$ upper limit is six orders of magnitude smaller than the latter one. We can therefore conclude that the agreement with the cosmological data leads to a massive gravity theory which has no impact on the dynamics of sytems in the low energy regime thus avoiding any difficulty in matching different scales.

Finally, since covariant massive gravity seems to be a consistent modification of GR, it would be interesting and worthwhile to further explore it. In particular, investigation should be improved by analyzing perturbed solutions and full prediction of cosmological perturbations.

\section*{Acknowledgements}

LP would like to thank R. Maartens for thoughtful advice. LP and NR would like to thank G. Vilasi for continuous encouraging and supporting. VFC is supported by Agenzia Spaziale Italiana (ASI) through contract Euclid\,-\,IC (I/031/10/0). NR and LP acknowledge partial support by a INFN/MICINN collaboration, Agenzia Spaziale Italiana (ASI) and the Italian Ministero Istruzione Universit\`{a} e Ricerca (MIUR) through the PRIN2008 project.


\begin{thebibliography}{99}

\bibitem{SNeIaNobel}
A.G. Riess et al.,  AJ, 116, 1009, 1998;
S. Perlmutter et al., ApJ, 517, 565, 1999;

\bibitem{others}
D.N. Spergel et al., ApJS, 148, 175, 2003;
S. Cole et al., MNRAS, 362, 505, 2005;
M. Kowalski et al., ApJ, 686, 749, 2008;
D.N. Spergel et al., ApJS, 170, 377, 2007

\bibitem{fierz39}
M. Fierz, W. Pauli, Proc. Roy. Soc. Lond. A 173, 211, 1939

\bibitem{derham10}
C. de Rham, G. Gabadadze, Phys. Rev. D, 82, 044020, 2010

\bibitem{chamseddine11}
A.H. Chamseddine, M.S. Volkov, Phys. Lett. B, 704, 652, 2011

\bibitem{boulware72}
D.G. Boulware, S. Deser, Phys. Rev. D, 6, 3368, 1972

\bibitem{arkani03}
N. Arkani\,-\,Hamed, H. Georgi, M.D. Schwartz, Annals Phys., 305, 96, 2005

\bibitem{creminelli05}
P. Creminelli, A. Nicolis, M. Papucci, E. Trincherini, JHEP, 0509, 003, 2005

\bibitem{derahm11}
C. de Rham, G. Gabadadze, A.J. Tolley, Phys. Rev. Lett., 106, 2311101, 2011

\bibitem{hassan11}
S.F. Hassan, R.A. Rosen, preprint arXiv\,:1106.3344, 2011

\bibitem{N11}
T.M. Nieuwenhuizen, Phys. Rev. D, 84, 024038, 2011

\bibitem{WMAP7}
E. Komatsu, K.M. Smith, J. Dunkley, C.L. Bennett, B. Gold, et al., ApJS, 192, 18, 2011

\bibitem{volkov}
M.S. Volkov, JHEP, 1201, 035 (2012)

\bibitem{Union2}
R. Amanullah, C. Lidman, C., D. Rubin, G. Aldering, P. Astier, et al. ApJ, 716, 712, 2010

\bibitem{Marcy}
V.F. Cardone, M. Perillo, S. Capozziello, MNRAS, 417, 1672, 2011

\bibitem{XS10}
L. Xiao, B.E. Schaefer, ApJ, 731, 103, 2011

\bibitem{JL02}
R. Jimenez, A. Loeb, ApJ, 573, 37, 2002

\bibitem{S10II}
D. Stern, R. Jimenez, L. Verde, S.A. Stanford, M. Kamionkowski, ApJS, 188, 280, 2010

\bibitem{S10I}
D. Stern, R. Jimenez, L. Verde, M. Kamionkowski, S.A. Stanford, JCAP, 02, 008, 2010

\bibitem{BAOth}
D.J. Eisenstein, W. Hu, M. Tegmark, ApJ, 504, 57, 1998;
A. Cooray, W. Hu, D. Huterer, M. Joffre, ApJ, 557, L7, 2001;
C.A. Blake, K. Glazebrook, ApJ, 594, 665, 2003;
W. Hu, Z. Haiman, Phys. Rev. D, 68, 063004, 2003

\bibitem{Eis05}
D.J. Eisenstein, I. Zehavi, D.W. Hogg, R. Scoccimarro, M.R. Blanton, et al., ApJ, 633, 560, 2005

\bibitem{BAOobs}
S. Cole, W.J. Percival, J.A. Peacock, P. Norberg, C.M. Baugh, MNRAS, 362, 505, 2005;
G. Hutsi, A\&A, 449, 891, 2006;
A.G. S\'anchez, M. Crocce, A. Cabr\'e, C.M. Baugh, E. Gazta$\tilde{{\rm n}}$aga, MNRAS, 400, 1643, 2009;
E.A. Kazin, M.R. Blanton, R. Scoccimarro, C.K. McBride, A.A. Berlind, et al., ApJ, 710, 1444, 2010;
C. Blake, A. Collister, S. Bridle, O. Lahav, MNRAS, 374, 1527, 2007;
N. Padmanabhan, D.J. Schlegel, U. Seljak, A. Makarov, N.A. Bahcall, et al., MNRAS, 378, 852, 2007;
M. Crocce, E. Gazta$\tilde{{\rm n}}$aga, A. Cabr\'e, A. Carnero, E. S\'anchez, MNRAS, 417, 2577, 2011

\bibitem{6dFGRS}
F. Beutler, C. Blake, M. Colless, D.H. Jones, L. Staveley\,-\,Smith, et al., MNRAS, 416, 3017, 2011

\bibitem{P10}
W.J. Percival, B.A. Reid, D.J. Eisenstein, N.A. Bahcall, T. Budavari, et al., MNRAS, 401, 2148, 2010

\bibitem{WiggleZ}
C. Blake, E.A. Kazin, F. Beutler, T.M. Davis, D. Parkinson, et al., MNRAS, 418, 1707, 2011

\bibitem{ShiftPar}
J.R. Bond, G. Efstathiou, M. Tegmark, MNRAS, 291, L33, 1997;
L. Page, M.R. Nolta, C., Barnes, C.L. Bennett, M. Halpern, et al. ApJS, 148, 233, 2003

\bibitem{HS96}
W. Hu, N. Sugiyama, ApJ, 471, 542, 1996

\bibitem{SHOES}
A. Riess, L. Macri, S. Casertano, M. Sosey, H. Lampeitl, et al., ApJ, 699, 539, 2009

\bibitem{K11}
K. Koyama, G. Niz, G. Tasinato, Phys. Rev. D, 84, 064033, 2011

\bibitem{GR92}
A. Gelman, D.B. Rubin, Stat. Sci., 7, 457, 1992

\bibitem{deffayet05}
C. Deffayet and J. W. Rombouts, Phys. Rev. D 72, 044003 (2005)

\bibitem{T88}
C. Talmadge, J.P. Berthias, R.W. Hellings, E.M. Standish, Phys. Rev. Lett., 61, 1988

\bibitem{disk}
M.E.S. Alves, O.D. Miranda, J.C.N. de Araujo, Gen. Rel. Grav., 39, 777,  2007

\bibitem{koyama12}
K. Koyama, G. Niz and G. Tasinato, Phys. Rev. D 84, 064033, 2011

\end{thebibliography}
\end{document}